# Stability of Scheduled Multi-access Communication over Quasi-static Flat Fading Channels with Random Coding and Joint Maximum Likelihood Decoding

KCV Kalyanarama Sesha Sayee, Utpal Mukherji
Dept. of Electrical Communication Engineering
Indian Institute of Science, Bangalore-560012, India
Email: sayee, utpal@ece.iisc.ernet.in

*Abstract*— We consider stability of scheduled multiaccess message communication with random coding and joint maximum-likelihood decoding of messages. The framework we consider here models both the random message arrivals and the subsequent reliable communication by suitably combining techniques from queueing theory and information theory. The number of messages that may be scheduled for simultaneous transmission is limited to a given maximum value, and the channels from transmitters to receiver are quasi-static, flat, and have independent fades. Requests for message transmissions are assumed to arrive according to an i.i.d. arrival process. Then, (i) we derive an outer bound to the region of message arrival rate vectors achievable by the class of stationary scheduling policies, (ii) we show for any message arrival rate vector that satisfies the outerbound, that there exists a stationary state-independent policy that results in a stable system for the corresponding message arrival process, and (iii) in the limit of large message lengths, we show that the stability region of message nat arrival rate vectors has information-theoretic capacity region interpretation.

## I. INTRODUCTION

Multi-access random-coded communication with independent decoding, of messages that arrive in a Poisson process to an infinite transmitter population, and that achieves any desired value for the random coding upper bound expected message error probability, by determining message signal durations appropriately, has been considered in [1] and [2]. Recently, in [3], a generalization and extension of the model in [1] and [2] was considered and the following assertions were proved: (i) in the limit of large message alphabet size, the stability region has an interference limited information-theoretic capacity interpretation, (ii) state-independent scheduling policies achieve this asymptotic stability region, and (iii) in the asymptotic limit corresponding to immediate access, the stability region for non-idling scheduling policies is identical irrespective of received signal powers. In independent decoding, each user is decoded *independently*, treating all other users as interference. Since independent decoding is suboptimal, we consider in the present work joint decoding of all user signals and establish results that are similar to the results shown in [3]. Some previous work with joint decoding is reported in [4].

In this paper we consider message (packet) communication from $J \geq 2$ transmitters to a receiver over a flat bandpass AWGN channel. Requests for message transmissions at different transmitters are generated in i.i.d. processes. Messages at transmitter-$j$, $1 \leq j \leq J$, are chosen from the message alphabet $\mathcal{M}_j$ consisting of $M_j \geq 2$ alternatives. Signals, representing messages, are to be communicated reliably; reliability is quantified by the tolerable joint message decoding error probability $p_e$. We assume that the receiver schedules messages for simultaneous transmission, i.e., the receiver can choose some numbers of messages from each of the $J$ transmitters. Due to the complexity involved in joint maximum likehood decoding of an arbitrary number of messages, the receiver is restricted to schedule at most a finite $\mathsf{K} \geq 1$ of messages at a time. This restriction gives rise to a set of possible schedules $\mathcal{S}_\mathsf{K}$, defined in Section II. The channels from transmitters to receiver are quasi-static, flat, and have independent fades. The actual communication is accomplished as follows. For a schedule $s \in \mathcal{S}_\mathsf{K}$ chosen by the receiver, the transmitters map their respective messages to codewords (signals) of length $N(s)$ and then transmit the signals. The length of the code word is carefully chosen so that reliable communication, quantified by $p_e$, is achieved.

The contributions in this paper are as follows. We derive an outer bound $\mathcal{R}_{out}$ to the stability region of message arrival rate vectors $\mathbb{E}A = (\mathbb{E}A_1, \mathbb{E}A_2, \ldots, \mathbb{E}A_J)$ achievable by the class of stationary scheduling policies. Next, we propose a class of stationary policies, called "*state-independent*" scheduling policies and denoted by $\Omega^\mathsf{K}$, and then characterize the stability region $\mathcal{R}(\omega)$ of message arrival rate vectors $\mathbb{E}A = (\mathbb{E}A_1, \mathbb{E}A_2, \ldots, \mathbb{E}A_J)$ achievable by any policy $\omega \in \Omega^\mathsf{K}$. We then go on to establish that for any message arrival processes with rate vector within the outerbound derived for stationary policies , there exists a state-independent scheduling policy $\omega$ such that the message system is stable. Finally, for a given set of average power constraints at the respective transmitters, we give information-theoretic capacity region interpretation to the stability region of message nat arrival rate vectors achievable by fixed schedules $s$

## II. THE INFORMATION THEORETIC MODEL

In this section we briefly touch upon the information-theoretic model of multiaccess communication, and discuss a random coding bound achievable by joint maximum likelihood decoding as derived in [5]. Let there be $J$ independent sources of information communicating to a receiver over a memoryless Gaussian channel. Let source-$j$'s alphabet be defined by $M_j$ possible message values and let $m_j$ and $\hat{m}_j$ denote the $j$th source output and its estimate at the receiver. Let $S$ denote any non-empty subset of the set of sources $\mathcal{J} = \{1, 2, \ldots, J\}$ and $\mathcal{P}(\mathcal{J})$ denote the set of all *non-empty* subsets of the set $\mathcal{J}$. For a given $S \in \mathcal{P}(\mathcal{J})$, we define an error event to be of *type-S* if the decoded joint message $\hat{m} = (\hat{m}_1, \hat{m}_2, \ldots, \hat{m}_J)$ and the original joint message $m = (m_1, m_2, \ldots, m_J)$ satisfy: $\hat{m}_j \neq m_j$ for $j \in S$ and $\hat{m}_j = m_j$ for $j \in S^c$. Assuming each source is encoded independently using block encoding, let $\bar{p}_{e,S}$ be the expected probability of a type-$S$ event over the ensemble of block codes; obviously the expected probability of error $\bar{p}_e = \sum_S \bar{p}_{e,S}$. We state here a random coding bound (Theorem 2 in [5]) on the expected error probability in decoding the joint message $m$ from the outputs of the first $N$ channel uses, for an AWGN channel, when the number of sources $J > 2$ and when specialized to Gaussian encoding at the respective sources.

*Theorem 2.1:* The expected error probability in decoding the joint message $m$ using joint maximum-likelihood decoding $\bar{p}_e = \sum_{S \in \mathcal{P}(\mathcal{J})} \bar{p}_{e,S}$, where $\forall \rho, 0 \leq \rho \leq 1$,

$$\bar{p}_{e,S} \leq \exp\left(\rho \sum_{j \in S} \ln M_j - N E_{o,S}\right)$$

$$\text{for} \quad E_{o,S} = \rho \ln\left(1 + \frac{\sum_{j \in S} P_j}{(1+\rho)\sigma^2}\right),$$

where $P_j$ is the average power assigned to the $j$th transmitter and $\sigma^2$ is the noise variance of AWGN channel. ∎

For future reference, we denote the random coding upper bound on the expected joint message decoding error probability as $\chi(\mathcal{J}, N) = \sum_{S \in \mathcal{P}(\mathcal{J})} \exp\left(\rho \sum_{j \in S} \ln M_j - N E_{o,S}\right)$. For a particular choice of $\rho$ and tolerable joint message decoding error probability $p_e$, let $N$ be smallest positive integer such that $\chi(\mathcal{J}, N) \leq p_e$. Then $\bar{p}_e \leq p_e$. Since

$$\begin{aligned}
p(\hat{m}_j \neq m_j) &= \sum_{S \in \mathcal{P}(\mathcal{J}): j \in S} \bar{p}_{e,S} \\
&< \sum_{S \in \mathcal{P}(\mathcal{J})} \bar{p}_{e,S} \\
&\leq \chi(\mathcal{J}, N),
\end{aligned}$$

the random coding upper bound on expected *joint message* decoding error probability also serves as an upper bound on the expected decoding error probability of any *individual message*.

In the following Lemma 2.1 we assert the following simple observation: for a given tolerable probability of joint message decoding error $p_e$, let $N$ be the smallest positive integer such that $\chi(\mathcal{J}, N) \leq p_e$. Suppose that *only* users in the set $S \in \mathcal{P}(\mathcal{J})$ are to be scheduled for transmission. Then, users in the set $S$ need code words of at most length $N$ to achieve the same decoding error probability $p_e$.

*Lemma 2.1:* Let $N$ be the smallest positive integer such that $\chi(\mathcal{J}, N) \leq p_e$, and let $S \in \mathcal{P}(\mathcal{J})$. Then

$$\chi(S, N) \leq \chi(\mathcal{J}, N)$$

∎

Since no closed form expression exists for $N$, we state an upper bound and a lower bound to $N$ in Lemma 2.2. We introduce the notation that, for any $x > 0$ and $q > 0$, $\lceil x \rceil_q = \min(n \geq 1 : x \leq nq)q$.

*Lemma 2.2:* For a given tolerable joint decoding error probability $p_e$, let $N$ be the smallest positive integer such that $\chi(\mathcal{J}, N) \leq p_e$. Then

(a) $N$ can be bounded as

$$N \geq \max_{S \in \mathcal{P}(\mathcal{J})} \frac{\left\lceil -\ln p_e + \rho \sum_{j \in S} \ln M_j \right\rceil_{E_{o,S}}}{E_{o,S}}$$

and

$$N \leq \max_{S \in \mathcal{P}(\mathcal{J})} \frac{\left\lceil -\ln \frac{p_e}{2^{\mathsf{K}}-1} + \rho \sum_{j \in S} \ln M_j \right\rceil_{E_{o,S}}}{E_{o,S}}$$

(b) for $1 \leq j \leq J$ and an integer $M \geq 2$, let $M_j = M$ and $\#S$ denote cardinality of the set $S$. Then

$$\lim_{M \to \infty} \frac{\ln M}{N} = \min_{S \in \mathcal{P}(\mathcal{J})} \frac{E_{o,S}}{\rho(\#S)}$$

∎

In what follows we allow for the possibility of scheduling *multiple* messages from a user. Let $s = (s_1, s_2, \ldots, s_J) \in \mathbb{Z}_+^J$ be a vector of non-negative integers and define the set $\mathcal{S}_{\mathsf{K}} = \left\{ s : 0 \leq \sum_{j=1}^J s_j \leq \mathsf{K} \right\}$, where $\mathcal{S}_{\mathsf{K}}$ denotes the set of all schedules that schedule at most $\mathsf{K}$ messages for simultaneous transmission. We assume that, for the schedule $s$, $j$th user encoder does joint encoding of $s_j$ messages. To interpret Theorem 2.1, Lemma 2.1 and 2.2 for the schedule $s \in \mathcal{S}_{\mathsf{K}}$, it is convenient to view the schedule $s$ as defining new message alphabets for the sources that are product versions of their original message alphabets. For example, for source-$j$ and for the schedule $s$, this product message alphabet is the Cartesian product of $s_j$ copies of the original message alphabet $\mathcal{M}_j$; hence the product message alphabet consists of $M_j^{s_j}$ different tuples of length $s_j$.

## III. QUEUEING-THEORETIC MODEL

In this section we derive a queueing-theoretic model for a $J$ user multiaccess message communication scheme, when requests for message transmission are randomly generated. This queueing model consists of $J$ queues, one for each source, and a single server whose service statistics depend on the state of the queues through the chosen scheduling policy.

Let maximum-likelihood decoding be used to decode the received word. Consider a fixed schedule $s$ and suppose that the tolerable message decoding error probability $p_e$ is given. The definition of service requirement that we consider for any source is the smallest positive integer $N(s)$ (length of the code word that each source transmits) such that $\chi(s, N(s)) \leq p_e$. For the schedule $s$, we say that source-$j$ receives a service quantum equivalent to $s_j$ units/slot; the total service quantum then is $\sum_{j=1}^{J} s_j$ units/slot. After receiving signal transmission over $N$ channel uses, the receiver will decode the joint message (schedule). A few remarks on the definitions of service requirement and service quantum are in order. The service requirement of a message depends on the schedule of which the message is a component message. In other words, a message by itself *cannot* characterize service requirement for itself unless it is the only message to constitute the schedule. The amount of service quantum available to each queue depends on the schedule.

Requests for message transmission are assumed to arrive at slot boundaries in batches. Let the random variable $A_j$, with finite moments $\mathbb{E}A_j$ and $\mathbb{E}A_j^2$, represent the number of messages that arrive in any slot at the $j$th queue, with the pmf $\Pr(A_j = k) = p_j(k)$, $k \geq 0$. We assume that $\{A_j\}$ are independent random variables. Let $\mathbb{E}A = (\mathbb{E}A_1, \mathbb{E}A_2, \ldots, \mathbb{E}A_J) \in \mathbb{R}_+^J$. Let $\lambda_j$ denote the arrival rate at queue-$j$. For channel bandwidth $W$, since each slot is of duration $\frac{1}{W}$, we have $\lambda_j = W\mathbb{E}A_j$.

Having defined service requirement and service quantum for a source and modelled message arrival processes, we are now in a position to analyze this message communication scheme with random coding and joint maximum-likelihood decoding when requests for message transmission arrive at random times. We construct a discrete-time countable state space Markov-chain model of this communication system and then analyze for the stability ($c$-regularity [6]) of the model. The stability analysis consists of characterizing the stability region $\mathcal{R}(\omega) \in \mathbb{R}_+^J$ of message arrival rate vectors $\mathbb{E}A$ for each policy $\omega$ in a class of stationary i"state-independent" scheduling policies, by obtaining appropriate drift conditions for suitably defined Lyapunov functions of the state of the Markov chain. In particular, we prove that the Markov chain is $c$-regular by applying Theorem 10.3 from [6], and then show finiteness of the stationary mean number of messages in the system.

## IV. A GENERAL OUTER BOUND TO THE STABILITY REGION

In this section, we derive an outerbound to the region of message arrival rate vectors $\mathbb{E}A$ for which the Markov-chain model is positive recurrent and has finite stationary mean for the number of messages, for the class of stationary scheduling policies. Later, in Section V, we propose a class of stationary scheduling policies, called "state-independent" scheduling policies and denoted by $\Omega^{\mathsf{K}}$, and then prove that for any message arrival processes $\{A_j; 1 \leq j \leq J\}$ with $\mathbb{E}A_j$ inside the outerbound, there exists a scheduling policy $\omega \in \Omega^{\mathsf{K}}$ such that the Markov-chain model is positive recurrent and has finite stationary mean for the number of messages.

Consider message arrival processes $\{A_j; 1 \leq j \leq J\}$ and a stationary scheduling policy $\omega$ that schedules at most $\mathsf{K}$ messages for a joint message transmission. Let $\pi_{\mathsf{K}}(s)$ be a probability measure on $\mathcal{S}_{\mathsf{K}}$. Define

$$\Psi_j = \sum_{\{s \in \mathcal{S}_{\mathsf{K}}: s_j > 0\}} \pi_{\mathsf{K}}(s) \frac{s_j}{N(s)}$$

and the set

$$\mathcal{R}_{out} = \bigcup_{\pi_{\mathsf{K}}(s)} \left\{ \beta \in \mathbb{R}_+^J : \beta_j \leq \Psi_j \right\} \quad (1)$$

*Theorem 4.1:* Let the Markov chain $\{X_n, n \geq 0\}$ be positive recurrent and yield finite stationary mean for the number of messages in the system for the message arrival processes $\{A_j\}$ and the stationary scheduling policy $\omega$. Then $\mathbb{E}A \in \mathcal{R}_{out}$. ∎

## V. STABILITY FOR STATE-INDEPENDENT SCHEDULING POLICIES

In this section we define the class of state-independent scheduling policies $\Omega^{\mathsf{K}}$, and then prove positive recurrence and finiteness of the stationary mean for the number of messages of the Markov-chain model for this class of scheduling policies. Formally, a policy in this class is defined by (i) a probability measure $\{p(s); s \in \mathcal{S}_{\mathsf{K}}\}$, and (ii) the mapping [1] $\{\omega : \mathcal{X} \times \mathcal{S}_{\mathsf{K}} \to \mathcal{S}_{\mathsf{K}}\}$. To implement a scheduling policy $\omega$ in $\Omega^{\mathsf{K}}$, we first classify message requests at any queue based on the particular schedule $s$ to be assigned to them.

For each message arrival at queue-$j$, a schedule $s \in \{s \in \mathcal{S}_{\mathsf{K}} : s_j > 0\}$ is chosen randomly with the fixed probability measure defined later in (3) and the message is further classified by assigning the class-$(j, s)$ to it. With this classification a message of class-$(j, s)$ will be scheduled to transmit only when the schedule $s$ gets chosen for transmission. One consequence of class sub-classification is that messages of class-$(j, s)$ will be required to use code words of length $N(s)$ for transmission, i.e., service requirement gets fixed. We first fix a scheduling policy $\omega = p(s)$ and then, in each time slot, a schedule $s$ is chosen from the set $\mathcal{S}_{\mathsf{K}}$, *independent* of the state $\alpha$, with probability $p(s)$. We constrain the operation of the system by requiring that there can be at most one on-going transmission [2] for any given schedule. For any policy $\omega \in \Omega^{\mathsf{K}}$, the queueing model consists of a number of message queues, one for each class-$(j, s)$. To define the state of the system, we keep track of the following information about each message class: for the message class-$(j, s)$, let $n_{js}(\alpha)$ denote the number of fresh [3] messages, $x_{js}$ the number of messages that are part of the on-going transmission, and $t_{js}$ the number of time-slots of transmission remaining for the on-going transmission to be completed. Define $\alpha_{js} = (n_{js}(\alpha), x_{js}, t_{js})$, the state information corresponding to message class-$(j,s)$ and then

$$\alpha = (\alpha_{js}; 1 \leq j \leq J, s \in \mathcal{S}_{\mathsf{K}}), \quad (2)$$

the state of the system.

---

[1] Suppose that schedule $s$ is chosen in state $\alpha$ with probability $p(s)$. Then the actual schedule that gets implemented is $s' = \omega(\alpha, s) \in \mathcal{S}_{\mathsf{K}}$ and is defined as follows. For $1 \leq j \leq J$, $s'_j = 0$ if $s_j = 0$. Let, for at least one message class-$(j, s)$ associated with the schedule $s$, $x_{js} \neq 0$. Then $s'_j = x_{js}$. Otherwise $s'_j = \min\{n_{js}(\alpha), s_j\}$

[2] A joint message for which at least one time-slot of transmission is complete and transmission for at least one more time-slot remains to be completed.

[3] We say that a message request is *fresh* if that message has not yet been scheduled for the first time, i.e., first code symbol of the corresponding code word is yet to be transmitted.

Now we discuss implementation of the scheduling policy, $\omega$. Suppose that the system is in state $\alpha$. Then the schedule to be selected for implementation in state $\alpha$ is a random variable and takes values in $\mathcal{S}_{\mathsf{K}}$. When trying to implement a schedule $s$ the following possibilities can occur:

1) For all of the message classes associated with the schedule $s$, there are no fresh messages present in the system; nor is there an ongoing transmission of schedule $s$. Then, no messages are scheduled in that state, and the system moves to next state as determined by the message arrival processes.
2) *No* on-going transmission of schedule $s$ is present in the system, and for at least one message class associated with the schedule $s$ there is at least one fresh message available. Then, a new joint message of schedule $s$ is scheduled, formed out of the fresh messages available with as many fresh messages of pertinent classes as are possible but not exceeding the respective maximum numbers specified by the schedule $s$.
3) There is an on-going transmission of schedule $s$ present in the system. Then that transmission is scheduled in that slot.

$\mathcal{X}$ is the countable set of state vectors $\alpha$ defined in (2). Let $V(\alpha)$ be a Lyapunov function defined on $\mathcal{X}$ and let $\mathcal{R}(\omega)$ denote the set of message arrival rate vectors $\mathbb{E}A$ for which the Markov chain $\{X_n, n \geq 0\}$ for the scheduling policy $\omega$ is positive recurrent and yields finite stationary mean for the number of messages of each class. Then we prove the following Lemma and two Theorems.

*Lemma 5.1:* For $\alpha \in \mathcal{X}$ and for message class-$(j,s)$, define $c_{js}(\alpha) = N(s)n_{js}(\alpha) + s_j t_{js}$. Next, define $c(\alpha) = 1 + \sum_{js} c_{js}(\alpha)$ and

$$V(\alpha) = \sum_{js} \frac{c_{js}^2(\alpha)}{2\left(p(s)s_j - \mathbb{E}A_{js}N(s)\right)}.$$

Then, for the scheduling policy $\omega$, the Markov chain is $c$-regular if, for each message class-$(j,s)$, $\mathbb{E}A_{js}N(s) < p(s)s_j$. ∎

*Theorem 5.1:* Let, for at least one message class-$(j,s)$, $\mathbb{E}A_{js} > \frac{p(s)s_j}{N(s)}$. Then the Markov-chain $\{X_n, n \geq 0\}$ is transient. ∎

To prove Theorem 5.1, we show that for the Lyapunov function $V(\alpha) = 1 - \theta^{N(s)n_{js}(\alpha) + x_{js}(\alpha)t_{js}}$, there exists a value for $\theta$, $0 < \theta < 1$, for which $V(\alpha)$ satisfies the conditions for the theorem for transience [7].

Define $\mu_j = (\mu_{js}, s \in \mathcal{S}_{\mathsf{K}} : s_j > 0)$ be a splitting probability vector defined by

$$\mu_{js} = \frac{\frac{p(s)s_j}{N(s)}}{\sum_{\{s' \in \mathcal{S}_{\mathsf{K}} : s'_j > 0\}} \frac{p(s')s'_j}{N(s')}}. \quad (3)$$

Then, given that a message arrives at queue-$j$, $\mu_{js}$ is the probability that the message request is assigned schedule $s$.

The sufficient condition for $c$-regularity of the Markov-chain $\{X_n, n \geq 0\}$ stated in Lemma 5.1 and the sufficient condition for transience stated in Theorem 5.1 together give the exact characterization of the stability region, as stated in the following theorem.

*Theorem 5.2:* For the scheduling policy $\omega$, the Markov chain $\{X_n, n \geq 0\}$ is

(a) positive recurrent and yields finite stationary mean for the number of messages, if, for each queue-$j$,

$$\mathbb{E}A_j < \sum_{\{s \in \mathcal{S}_{\mathsf{K}} : s_j > 0\}} \frac{p(s)s_j}{N(s)}, \quad \text{and}$$

(b) transient if, for at least one message class-$(j,s)$,

$$\mathbb{E}A_{js} > \frac{p(s)s_j}{N(s)}$$

∎

Define

$$\psi_j = \sum_{\{s \in \mathcal{S}_{\mathsf{K}} : s_j > 0\}} p(s) \frac{s_j}{N(s)}$$

and the set

$$\mathcal{R}\left(\Omega^{\mathsf{K}}\right) = \bigcup_{p(s) \in \Omega^{\mathsf{K}}} \left\{\beta \in \mathbb{R}_+^J : \beta_j < \psi_j\right\} \quad (4)$$

*Corollary 5.1:* For any given message arrival rate vector $\mathbb{E}A \in \mathcal{R}\left(\Omega^{\mathsf{K}}\right)$ there exists a scheduling policy $p(s) \in \Omega^{\mathsf{K}}$ such that the Markov chain is positive recurrent and yields finite stationary mean for the number of messages of each class. ∎

From (1) and (4), we note [4] that $\mathcal{R}\left(\Omega^{\mathsf{K}}\right) = \mathcal{R}_{out}^o$. This observation essentially states that, if a stationary scheduling policy is stable for the message arrival processes $\{A_j\}$, then there exists a state-independent scheduling policy which makes the Markov-chain stable for the same message arrival processes $\{A_j\}$.

*Proof:* Suppose that, for some stationary scheduling policy, the Markov-chain model $\{X_n, n \geq 0\}$ is stable for the message arrival processes $\{A_j; 1 \leq j \leq J\}$. Let $\pi_{\mathsf{K}}(s)$ be the induced stationary probability distribution on the set of schedules $\mathcal{S}_{\mathsf{K}}$. Let $\pi_{\mathsf{K}}(0) > 0$ be the stationary probability that *no* schedule is served in a time-slot. In the steady state, let $\mathbb{E}A_{js}$ be the rate at which joint messages of composition $s$ finish service requirement. Then $\mathbb{E}A_{js}N(s) = \pi_{\mathsf{K}}(s)s_j$.

Let us define a new probability distribution $\{p(s), \mathcal{S}_{\mathsf{K}}\}$ as follows: for any non-empty schedule $s \in \mathcal{S}_{\mathsf{K}}$, define $p(s) = \pi_K(s) + \epsilon_s$ such that $\sum_s \epsilon_s = \pi_K(0)$, $\epsilon_s > 0$. Let a state-independent scheduling policy $\omega$ be defined by $\omega = \{p(s), \mathcal{S}_K\}$. Then, for any class-$(j,s)$, $\mathbb{E}A_{js}N(s) < p(s)s_j$. That is, for the message arrival processes $\{A_j, 1 \le j \le J\}$, the state-independent policy $\omega = \{p(s), \mathcal{S}_K\}$ makes the Markov-chain stable. ∎

---

[4]Interior of the set $A$ is denoted by $A^o$.

## VI. INFORMATION-THEORETIC CAPACITY INTERPRETATION

In this section we give information-theoretic capacity interpretation to the stability region of message nat arrival rate vectors $\mathbb{E}\tilde{A}$. A formal statement of this interpretation is made in Theorem 6.1. Let $\tilde{A}_j = A_j \ln M_j$ denote the nat arrival random variable corresponding to message class-$j$. Then, the message system for the fixed schedule $s$ is stable for nat arrival rates satisfying the following inequality: for each class $j$,

$$\mathbb{E}\tilde{A}_j < \frac{s_j \ln M_j}{N(s)}. \quad (5)$$

Inequality (5) follows trivially from Theorem 5.2. We can observe that the quantity $\frac{s_j \ln M_j}{N(s)}$ denotes the threshold coding rate for message class-$j$ under the schedule $s$. For $1 \le j \le J$ and integer $M \ge 2$, let $M_j = M$. We are interested in evaluating inequality (5) in the limit of $M \to \infty$ and $\rho \to 0$. Define $\overline{R}_j(s) = \lim_{\rho \to 0} \lim_{M \to \infty} \frac{s_j \ln M}{N(s)}$, $\overline{R}(s) = (\overline{R}_j(s), 1 \le j \le J)$, and the hypercube $\overline{\mathcal{R}}(s) \in \mathbb{R}_+^J$ defined by $\overline{R}(s)$. Then $\overline{R}_j(s)$ is determined using part $(b)$ of Lemma 2.2 and its application gives the following:

$$\overline{R}_j(s) = \lim_{\rho \to 0} \min_{S \in \mathcal{P}(\mathcal{J})} \frac{s_j E_{o,S}}{\rho \sum_{k \in S} s_k}$$

When messages from each queue are encoded jointly into Gaussian code words of power as determined by their message class, the expression $E_{o,S}$ reduces to $\rho \ln\left(1 + \frac{\sum_{j \in S} P_j}{(1+\rho)\sigma^2}\right)$. Thus, for Gaussian encoding

$$\overline{R}_j(s) = \min_{S \in \mathcal{P}(\mathcal{J})} \frac{s_j \ln\left(1 + \frac{\sum_{j \in S} P_j}{\sigma^2}\right)}{\sum_{k \in S} s_k} \quad (6)$$

Consider a Gaussian multiple access system with $J$ independent sources with powers $P = (P_1, P_2, \ldots, P_J)$ and the noise variance $\sigma^2$. Then, the capacity region $\mathcal{C}(P, \sigma^2) \subset \mathbb{R}_+^J$ is characterized as follows. Define $r = (r_1, r_2, \ldots, r_J) \in \mathbb{R}_+^J$. Then,

$$\mathcal{C}(P,\sigma^2) = \left\{ r : \sum_{k \in S} r_k \le \ln\left(1 + \frac{\sum_{k \in S} P_k}{\sigma^2}\right), S \in \mathcal{P}(\{1,2,\ldots,J\}) \right\}$$

Let $\mathcal{C}^o(P, \sigma^2)$ represent the interior of $\mathcal{C}(P, \sigma^2)$.

We show in the following Theorem 6.1 that, for any rate vector $R \in \mathcal{C}^o(P, \sigma^2)$, there exists a schedule $s$ such that the Markov chain $\{X_n, n \ge 0\}$, for schedule $s$ and arrival processes $\{\tilde{A}_j\}$ with $\mathbb{E}\tilde{A} = r$, is stable. That is, the achievable asymptotic stable region of message nat arrival rate vectors and the interior of the capacity region $\mathcal{C}(P, \sigma^2)$ are the same.

*Theorem 6.1 (Capacity Interpretation):*

$$\bigcup_{K \ge 1} \bigcup_{\{s \in \mathcal{S}_K\}} \overline{\mathcal{R}}(s) = \mathcal{C}^o(P, \sigma^2)$$

*Proof:* We first show that $\mathcal{C}^o(P, \sigma^2) \subset \cup_{K \ge 1} \cup_{\{s \in \mathcal{S}_K\}} \overline{\mathcal{R}}(s)$. Let $r = (r_1, r_2, \ldots, r_J) \in \mathcal{C}^o(P, \sigma^2)$. There exists an arbitrarily small $\epsilon > 0$ such that $r + \epsilon = (r_1 + \epsilon, r_2 + \epsilon, \ldots, r_J + \epsilon) \in \mathcal{C}^o(P, \sigma^2)$. Consider a schedule $s$ such that, for each $j$,

$$s_j \propto r_j + \epsilon, \text{ and for } i \ne j, \frac{s_i}{s_j} = \frac{r_i + \epsilon}{r_j + \epsilon}$$

Now, with $s_j$ chosen as suggested, it can be shown that the asymptotic coding rate for message class-$j$, $\overline{R}_j(s) > r_j + \epsilon$. Since $\sum_{k \in S} s_k < \ln\left(1 + \frac{\sum_{k \in S} P_k}{\sigma^2}\right)$ for each $S \in \mathcal{P}(\mathcal{J})$, we see from equation (6) that for every $S \in \mathcal{P}(\mathcal{J})$, $\frac{s_j}{\sum_{k \in S} s_k} \ln\left(1 + \frac{\sum_{k \in S} P_k}{\sigma^2}\right) > r_j + \epsilon$. That is, $r \in \overline{\mathcal{R}}(s)$ and since $s \in \mathcal{S}_K$ for some $K \ge 1$, we have $r \in \cup_{K \ge 1} \cup_{\{s \in \mathcal{S}_K\}} \overline{\mathcal{R}}(s)$.

Next, we show that $\cup_{K \ge 1} \cup_{\{s \in \mathcal{S}_K\}} \overline{\mathcal{R}}(s) \subset \mathcal{C}^o(P, \sigma^2)$. Let $R \in \cup_{K \ge 1} \cup_{\{s \in \mathcal{S}_K\}} \overline{\mathcal{R}}(s)$. Then, for some schedule $s$, $R \in \overline{\mathcal{R}}(s)$. Since the set $\mathcal{C}^o(P, \sigma^2)$ is characterized by $2^J - 1$ constraints, we show that $R$ satisfies all those $2^J - 1$ constraints. Let $S$ be a non-empty subset of the set $\{1, 2, \ldots, J\}$. For each $j \in S$, we have

$$R_j < \frac{s_j}{\sum_{k \in S} s_k} \ln\left(1 + \frac{\sum_{j \in S} P_j}{\sigma^2}\right)$$

Then $\sum_{k \in S} R_k < \ln\left(1 + \frac{\sum_{j \in S} P_j}{\sigma^2}\right)$. Since this is true for any non-empty subset $S$, we conclude that $R \in \mathcal{C}^o(P, \sigma^2)$. ∎

## REFERENCES

[1] I.E. Telatar and R.G. Gallager. "Combining Queueing Theory with Information Theory for Multiaccess". *IEEE Journal on Selected Areas in Communications*, 13(6):963–969, August 1995.
[2] I.E Telatar. *"Multi-access Communications with Decision Feedback Decoding"*. Ph.D. Dissertation, Report CICS-TH-33, Dept. of Electrical Engineering and Computer Science, Massachusetts Institute of Technology, 1992.
[3] KCV Kalyanarama Sesha Sayee and U. Mukherji. "Stability of Scheduled Multi-access Communication over Quasi-static Flat Fading Channels with Random Coding and Independent Decoding. *Proceedings, IEEE International Symposium on Information Theory*, pages 2261–2265, September 2005.
[4] I.E. Telatar. "Combining Queueing Theory with Information Theory for Multi-access with Joint Decoding". In *Proc. IEEE Information Theory Workshop*, page 51, St. Louis, April 1995.
[5] R. G. Gallager. "A Perspective on Multiaccess Channels". *IEEE Transactions on Information Theory*, IT-31(2):124–142, March 1985.
[6] S.P. Meyn. *"Stability, Performance Evaluation, and Optimization"*, chapter in Handbook of Markov Decision Processes (E.A. Feinberg and A. Shwartz, ed.). Kluwer Academic, 2002.
[7] S.P. Meyn and R.L. Tweedie. *Markov Chains and Stochastic Stability*. Springer Verlag, 1993.